\documentclass[12pt]{article}
\usepackage{graphicx}
\usepackage{epsfig}
\def\be {\begin{equation}}
\def\ee {\end{equation}}
\def\ba {\begin{eqnarray}}
\def\ea {\end{eqnarray}}

\def\bi {\begin{itemize}}
\def\ei {\end{itemize}}
\begin{document}
\def\bea{\begin{eqnarray}}
\def\eea{\end{eqnarray}}

\title{\textbf{Statefinder diagnostic and stability
of modified gravity consistent with holographic and new agegraphic
dark energy }}

\author{  \textbf{M. R. Setare} \thanks{
E-mail: rezakord@ipm.ir}\\{Department of Science of Bijar, University of Kurdistan, Bijar, Iran} \\
\textbf{ Mubasher Jamil} \thanks{E-mail: mjamil@camp.nust.edu.pk
 }\\
{ Center for Advanced Mathematics and Physics, National University
of }\\{Sciences and Technology, Islamabad, Pakistan}}
\maketitle

\begin{abstract}
\vspace*{1.5cm} Recently one of us derived the action of modified
gravity consistent with the holographic and new-agegraphic dark
energy. In this paper, we investigate the stability of the
Lagrangians of the modified gravity as discussed in [M. R.
Setare, Int. J. Mod. Phys. D 17 (2008) 2219; M. R. Setare,
Astrophys. Space Sci. 326 (2010) 27]. We also calculate the
statefinder parameters
which classify our dark energy model.

\end{abstract}

\maketitle
\newpage
\section{Introduction}
Nowadays it is strongly believed that the universe is experiencing
an accelerated expansion. Recent observations from type Ia
supernovae \cite{SN} in associated with Large Scale Structure
\cite{LSS} and Cosmic Microwave Background anisotropies \cite{CMB}
have provided main evidence for this cosmic acceleration. There
are two ways to explain the current accelerated expansion of the
universe.  The first one is to introduce some unknown matter,
which is called dark energy in the framework of general
relativity.

On the other hand the nature of dark energy is ambiguous. The
simplest candidate of dark energy is a cosmological constant with
the equation of state parameter $\omega=-1$. However, this scenario
suffers from serious problems like a huge fine tuning and the
coincidence problem \cite{4}. Alternative models of dark energy
suggest a dynamical form of dark energy, which is often realized by
one or two scalar fields. In this respect, dark energy components
such as quintessence \cite{quintessence}, K-essence \cite{kessence},
tachyon \cite{tachyon}, phantom \cite{10}, ghost condensate and
quintom \cite{quintom}, and so forth.

Although going beyond the above effective description requires a
deeper understanding of the underlying theory of quantum gravity
\cite{Witten:2000zk} unknown at present, physicists can still make
some attempts to probe the nature of dark energy according to some
basic quantum gravitational principles. An example of such a
paradigm is the holographic dark energy scenario, constructed in the
light of the holographic principle
\cite{Cohen:1998zx,Horava:2000tb,Hsu:2004ri,Li:2004rb}. Its
framework is the black hole thermodynamics \cite{BH22} and the
connection (known from AdS/CFT correspondence) of the UV cut-of of a
quantum field theory, which gives rise to the vacuum energy, with
the largest distance of the theory \cite{Cohen:1998zx}. Thus,
determining an appropriate quantity $L$ to serve as an IR cut-off,
imposing the constraint that the total vacuum energy in the
corresponding maximum volume must not be greater than the mass of a
black hole of the same size, and saturating the inequality, one
identifies the acquired vacuum energy as holographic dark energy:
\begin{equation}\label{de}
\rho_\Lambda=\frac{3c^2}{8\pi G L^2},
\end{equation}
with $G$ the Newton's gravitational constant and $c$ a constant. The
holographic dark energy scenario has been tested and constrained by
various astronomical observations
\cite{obs3a,obs2,obs1,Wu:2007fs,obs3} and it has been extended to
various frameworks \cite{nonflat,holoext,intde}.

More recently a new dark energy model, dubbed agegraphic dark energy
has been proposed. These models take into account the Heisenberg
uncertainty relation of quantum mechanics together with the
gravitational effect in general relativity. The agegraphic dark
energy models assume that the observed dark energy comes from the
spacetime and matter field fluctuations in the universe
\cite{Cai1,Wei2,Wei1}. Since in agegraphic dark energy model the age
of the universe is chosen as the length measure, instead of the
horizon distance, the causality problem in the holographic dark
energy is avoided. The agegraphic models of dark energy  have been
examined and constrained by various astronomical observations
\cite{age}.

Another alternative approach to explain the universe's late-time
acceleration is modifying the General Relativity itself \cite{30},
and in the simplest case replace $R$ with $f(R)$ in the action which
is well known as $f(R)$ gravity. Here $f(R)$ is an arbitrary
function of scalar curvature. Although there are some works with
related subjects on crossing of the phantom divide line in the
framework of modified gravity \cite{{c13},{c14}}, but Ref.\cite{c15}
was the first paper that has investigated a modified gravity model
realizing $\omega$ across -1. The authors of Ref.\cite{c15} have
shown an explicit model of modified gravity in which a crossing of
the phantom divide can occur and relation between scalar field
theories with property of $\omega$ crossing -1 and the corresponding
modified gravity theories have been investigated. Observationally
the possibility of phantom crossing was suggested by early supernova
data sets \cite{ref1} while the situation concerning present data is
rather ambiguous \cite{ref2}.

In paper \cite{setare1}, using the holographic model of dark energy
in spatially flat universe, one author of this paper has obtained
equation of state for holographic dark energy density in framework
of modified gravity for a universe enveloped by $R_h$ as the
system's IR cut-off. Also he has developed a reconstruction scheme
for the modified gravity with $f(R)$ action. He could
 to obtain a differential equation for $f(R)$, the solution of this
differential equation give us a modified gravity action which is
consistent with holographic dark energy scenario. In
\cite{setare2} this investigation extended to new agegraphic dark
energy model.

\section{Holographic dark energy}

The energy density of holographic dark energy is given by ($8\pi
G=1$)
\begin{equation}\label{hde}
\rho_D=3c^2R_h^{-2},
\end{equation}
where $c$ is a constant and $R_h$ is the future event horizon
specified by
\begin{equation}\label{Rh}
R_h\equiv
a(t)\int\limits_t^\infty\frac{dt}{a'}=a(t)\int\limits_{a(t)}^\infty\frac{da'}{Ha'^2}.
\end{equation}
The critical energy density $\rho_{cr}$ is given by the following
expression
\begin{equation}\label{rcr}
\rho_{cr}=3H^2,
\end{equation}
which helps us to write the dimensionless density parameter for dark
energy
\begin{equation}\label{Od}
\Omega_D=\frac{\rho_D}{\rho_{cr}}=\frac{c^2}{R_h^2H^2}.
\end{equation}
Differentiating (\ref{Rh}) and using (\ref{Od}), we are able to
write
\begin{equation}
\dot R_h=HR_h-1=\frac{c}{\sqrt{\Omega_D}}-1.
\end{equation}
The energy conservation equation for (any form of) dark energy is
\begin{equation}\label{dd}
\dot\rho_D+3H(1+\omega_D)\rho_D=0.
\end{equation}
Differentiating (\ref{hde}) w.r.t $t$ and using (\ref{Od}) yields
\begin{equation}\label{dd1}
\dot\rho_D=-2\frac{\rho_D}{R_h}\Big(\frac{c}{\sqrt{\Omega_D}}-1\Big).
\end{equation}
Using (\ref{dd1}) in (\ref{dd}) yields
\begin{equation}
\omega_{HDE}=-\frac{1}{3}-\frac{2}{3c}\sqrt{\Omega_D},
\end{equation}
the equation of state parameter for the holographic dark energy.

\section{Stability of modified gravity with HDE}

It is shown in \cite{setare1} that a modified gravity consistent
with the holographic dark energy in flat space has the following
form:
\begin{equation}\label{fhde}
f(R)=C_1R^{u}+C_2R^{v}+\frac{c^2(1-h_0)^2R}{2h_0^2},
\end{equation}
where $h_0$ is a constant which is related to the scale factor as
$a(t)=a_0(t_s-t)^{h_0}$, $t_s$ and $a_0$ are also constants. The
expressions of $u$, $v$, $C_1$ and $C_2$ are discussed in
\cite{setare1}.

For a phenomenological $f(R)$ model to be viable, it must meet a
certain list of viability conditions \cite{star}: \textit{classical
and quantum stability} which involves $f'(R)>0$ and $f''(R)>0$ (the
first positive derivative means that gravity is attractive and the
graviton is not a ghost while the second positive derivative is used
to avoid the Doglov-Kawasaki instability); stable Newtonian limit
for all values of $R$; absence of deviations from general relativity
and existence of a future de Sitter asymptote. We are interested
here to check the stability of our $f(R)$ models only. The stability
can be checked by calculating the double derivative of $f(R)$ in
(\ref{fhde}):
\begin{equation}
f''(R)\equiv\frac{d^2f}{dR^2}=C_1u(u-1)R^{u-2}+C_2v(v-1)R^{v-2}.
\end{equation}
Now $f''(R)>0$ if $C_1>0$, $C_2>0$, $u>1$ and $v>1$. This case
corresponds to stability while a negative double derivative results
in instability. We are not interested in the later case. There are
some suggestions that the instabilities associated with $f(R)$
action can be removed by adding quadratic and higher order terms
along with the Ricci scalar $R$ \cite{noji}. Thus due to abundance
of free parameters, the instabilities associated with $f(R)$ action
are naturally alleviated. Note that when $f''(R)=0$ for a suitable
choice of model parameters (e.g. $C_1=0=C_2$) then the theory
reduces to general relativity and is of no interest here.

\section{New agegraphic dark energy}

The energy density of the new agegraphic dark energy is given by
\begin{equation}\label{nade}
\rho_D=3n^2\eta^{-2},
\end{equation}
where $n$ is a constant and $\eta$ is the conformal time defined by
\begin{equation}
\eta=\int\frac{dt}{a}=\int\limits_0^a\frac{da'}{a'^2H}.
\end{equation}
Using (\ref{nade}) and (\ref{rcr}), we obtain
\begin{equation}
H\eta=\frac{n}{\sqrt{\Omega_D}}.
\end{equation}
Differentiating (\ref{nade}), we get
\begin{equation}\label{dd2}
\dot\rho_D=-\frac{2}{a\eta}\rho_D.
\end{equation}
Using (\ref{dd2}) in (\ref{dd}), we obtain
\begin{equation}\label{od1}
\omega_{NADE}=-1+\frac{2}{3an}\sqrt{\Omega_D}.
\end{equation}

\section{Stability of modified gravity with NADE}

It is shown in \cite{setare2} that a modified gravity consistent
with the agegraphic dark energy in flat space has the following
form:
\begin{equation}\label{f2}
f(R)=C_3R^{u}+C_4R^{v}+\frac{C_5}{R^{h_0-1}}.
\end{equation}
The expressions of $C_4$ and $C_5$ are discussed in \cite{setare2}.

Now stability can be checked by calculating the double derivative of
$f(R)$ in (\ref{f2}):
\begin{equation}
\frac{d^2f}{dR^2}=C_3u(u-1)R^{u-2}+C_4v(v-1)R^{v-2}+h_0(h_0-1)\frac{C_5}{R^{1+h_0}}.
\end{equation}
Now $f''(R)>0$ if $C_3>0$, $C_4>0$, $C_5>0$, $u>1$, $v>1$ and
$h_0>1$. This case corresponds to stability while a negative double
derivative results in instability. We are not interested in the
later case. It should be noted that the order of these coefficients
should of the order of $H_0^2\simeq 10^{-66}eV^2$, so that the
correction terms play their role at the present time to produce
cosmic acceleration. We would also mention that at present there is
no indication what should be the correct modification of the
Einstein's general relativity from the observations. On the other
hand, some recent astrophysical observations of weak lensing and
galaxy velocities indicate that general relativity is still the best
fit of the astrophysical data \cite{Reinabell}. The present
considered models contain numerous parameters and observational data
can constrain them only.

\section{Statefinder diagnostic}

In this section, we calculate the statefinder parameters for the
above two models of dark energy. Sahni et al \cite{sahni} introduced
a pair of cosmological diagnostic pair $\{r,s\}$ which they termed
as Statefinder. The two parameters are dimensionless and are
geometrical since they are derived from the cosmic scale factor
alone, though one can rewrite them in terms of the parameters of
dark energy. Additionally, the pair gives information about dark
energy in a model independent way i.e. it categorizes dark energy in
the context of background geometry only which is not dependent on
the theory of gravity. Hence geometrical variables are universal.
Also this pair generalizes the well-known geometrical parameters
like the Hubble parameter and the deceleration parameter. This pair
is algebraically related to the equation of state of dark energy and
its first time derivative.

The statefinder parameters $\{r,s\}$ are defined as
\cite{sahni,yifu}
\begin{equation}\label{rs}
r\equiv\frac{\stackrel{...}a}{aH^3},\ \
s\equiv\frac{r-1}{3(q-1/2)}.
\end{equation}
Note that in the derivation of the above parameters, a spatially
flat FRW spacetime is assumed. A useful alternative form of
(\ref{rs}) is
\begin{eqnarray}
r&=&1-\frac{3}{2}\Omega_D\Big[
\frac{\dot\omega_D}{H}-3\omega_D(1+\omega_D) \Big],\label{r}\\
s&=&\frac{-1}{3\omega_D}\Big[
\frac{\dot\omega_D}{H}-3\omega_D(1+\omega_D) \Big].\label{s}
\end{eqnarray}
One can immediately see that for cosmological constant with constant
equation of state ($\omega_D=-1$), we have $\{1,0\}$. Moreover
$\{1,1\}$ represents the standard cold dark matter model containing
no radiation while Einstein static universe corresponds to
$\{\infty,-\infty\}$ \cite{debnath}. In literature, the diagnostic
pair is analyzed for various dark energy candidates including
holographic dark energy \cite{zhang}, agegraphic dark energy
\cite{wei}, quintessence \cite{zhang1}, dilaton dark energy
\cite{dilaton}, Yang-Mills dark energy \cite{yang}, viscous dark
energy \cite{vis}, interacting dark energy \cite{pavon}, tachyon
\cite{shao}, modified Chaplygin gas \cite{debnath1}, $f(R)$ gravity
\cite{song} and dark energy model with variable constants
\cite{jamildeb} to name a few.

The dimensionless dark energy density parameter is
\cite{setare1,setare2}
\begin{equation}
\Omega_D=\frac{1}{3H^2}\Big[ f(R)-6(\dot H+H^2-H\frac{d}{dt})f'(R)
\Big],
\end{equation}
where prime denotes differentiation w.r.t $R$. The above expression
can be written as
\begin{equation}\label{Od1}
\Omega_D=\frac{1}{3H^2}\Big[ f(R)-6(\dot H+H^2)F(R)+6HF_{,R}\dot R
\Big],
\end{equation}
where we have used the  following relations
\begin{equation}
F(R)\equiv f'(R),\ \ \frac{d}{dt}f'(R)=f_{,RR}\dot R=F_{,R}\dot R.
\end{equation}
We proceed to calculate statefinder parameters for the NADE.
Substituting (\ref{Od1}) in (\ref{od1})
\begin{equation}
\omega_{NADE}=-1+\frac{2}{3\sqrt{3}anH}\sqrt{ f(R)-6(\dot
H+H^2)F(R)+6HF_{,R}\dot R}.
\end{equation}
We differentiate (\ref{od1}) w.r.t $t$ and obtain
\begin{equation}\label{d3d}
\dot{\omega}_{NADE}=\frac{2}{3na}H\sqrt{\Omega_D}\Big(
\frac{\dot{\Omega}_D}{2H\Omega_D} -1\Big).
\end{equation}
Differentiating (\ref{Od1}) leads to
\begin{eqnarray}\label{dod}
\dot{\Omega}_D&=&\frac{-2\dot H}{3H^3}\Big[  f(R)-6(\dot
H+H^2)F(R)+6HF_{,R}\dot R \Big]\nonumber\\&&+\frac{1}{3H^2}[F\dot
R-6(\stackrel{...}H+2H\dot H)F+6\dot HF_{,R}\dot R+6H(F_{,RR}\dot
{R}^2+F_{,R}\ddot R)].
\end{eqnarray}
Similarly, we obtain parameters for the HDE:
\begin{equation}
\omega_{HDE}=\frac{-1}{3}-\frac{2}{3\sqrt{3}cH}\sqrt{ f(R)-6(\dot
H+H^2)F(R)+6HF_{,R}\dot R}.
\end{equation}
\begin{equation}\label{123}
\dot{\omega}_{HDE}=\frac{-1}{3c}\frac{\dot{\Omega}_D}{\sqrt{\Omega_D}}.
\end{equation}
Using (\ref{od1}) and (\ref{d3d}) in (\ref{r}) and (\ref{s}), we
obtain
\begin{equation}\label{r1}
r_{NADE}=1-\frac{3}{an}\Omega_D^{3/2}\Big[
1-\frac{\sqrt{\Omega_D}}{an}+
\frac{\dot{\Omega}_D}{4H\Omega_D}\Big].
\end{equation}
\begin{equation}\label{s1}
s_{NADE}=\frac{2}{3an}\sqrt{\Omega_D}\Big[
1-\frac{-1+\frac{\dot{\Omega}_D}{2H\Omega_D}}{
-3+\frac{2}{an}\sqrt{\Omega_D}}\Big].
\end{equation}
Similarly we can obtain $\{r,s\}$ parameters for the HDE as
\begin{equation}\label{r2}
r_{HDE}=1-\frac{1}{2}\Omega_D\Big[
-\frac{1}{cH}\frac{\dot{\Omega}_D}{\sqrt{\Omega_D}}+2\Big(
1-\frac{4}{c^2}\Omega_D \Big) \Big].
\end{equation}
\begin{equation}\label{s2}
s_{HDE}=\frac{1}{3\Big( 1+\frac{2}{c} \sqrt{\Omega_D}\Big)}
\Big[-\frac{1}{cH}\frac{\dot{\Omega}_D}{\sqrt{\Omega_D}}+2\Big(
1+\frac{2}{c}\sqrt{\Omega_D} \Big)\Big( 1-\frac{1}{c}\sqrt{\Omega_D}
\Big) \Big].
\end{equation}

\section{Conclusion}

Besides compatibility with experimental data, the minimal criteria
that a modified gravity theory must satisfy in order to be viable
are \cite{faraoni}: reproducing the desired dynamics of the universe
including an inflationary era, followed by a radiation era and a
matter era and, finally, by the present acceleration epoch.
Moreover, the theory must have Newtonian and post-Newtonian limits
compatible with the available solar system observational data. And
the theory must be stable at the classical and quantum level.

In this paper, we have studied the stability issues associated with
the HDE and NADE in the framework of modified gravity. Our study
provides a more general framework to study such models. It should be
noted that the models of \cite{capo} is a special case of our paper
for the particular choices of parameters such as $C_3=0$ (or
$C_4=0$, but not both simultaneously) and $u=1(v=0)$ or $v=1(u=0)$
and $C_5=-\mu^4$. Also note that the suggestion of \cite{capo} is
ruled out since it was incompatible with the observational data of
solar system tests. A more appropriate $f(R)$ action is provided by
Setare \cite{setare1,setare2} and here we obtained constraints on
the parameters of our model if the $f(R)$ gravity theory is stable
i.e. $f''(R)>0$. The opposite case is not discussed for not general
interest and it corresponds to Dolgov-Kawasaki instability
\cite{faraoni}. Finally we calculated the statefinder diagnostic
parameters to characterize the present models of dark energy and
plotted their trajectories in Fig 1 and 2 for a suitable choice of
model parameters. The dots in both figures represent the LCDM model
for which $\{r,s\}=\{1,0\}$. We emphasize here that the behavior of
statefinder parameters in the figures is due to a selected choice of
model parameters while it can change for other values of parameters.

\subsubsection*{Acknowledgments}

We would like to thank the referee for giving very enlightening
comments to improve this work. MJ would like to thank the Abdus
Salam ICTP, Trieste, Italy for the local hospitality where part of
this work was completed.

\newpage

\begin{figure}
\includegraphics[scale=.9]{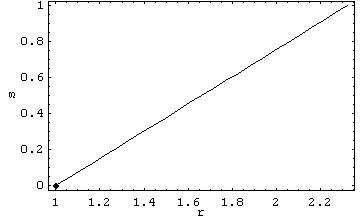}\\
\caption{The statefinder parameters for NADE are plotted for the
parametric values $h_0=1.1$, $n=0.03$, $C_3=0.02$, $C_4=0.03$,
$C_5=0.04$ and $u=v=1.1$}
\end{figure}

\begin{figure}
\includegraphics[scale=.9]{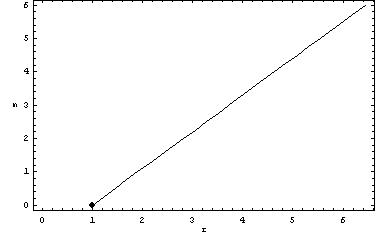}\\
\caption{The statefinder parameters for HDE are plotted for the
parametric values $h_0=0.9$, $c=0.5$, $C_1=0.02$, $C_2=0.3$,and
$u=1.1$ and $v=1.2$}
\end{figure}

\end{document}